\documentclass[12pt]{article}
\usepackage{amsmath,amssymb,amsfonts,epsf}
\usepackage[nosort]{cite}
\usepackage[margin=1in]{geometry}

\makeatletter \@addtoreset{equation}{section} \makeatother

\newcommand{\fft}[2]{{\frac{#1}{#2}}}
\newcommand{\ft}[2]{{\textstyle\frac{#1}{#2}}}

\def\nn{\nonumber}

\newcommand{\be}{\begin{equation}}
\newcommand{\ee}{\end{equation}}
\def\ba{\begin{array}}
\def\ea{\end{array}}
\def\ft#1#2{{\textstyle{\frac{\scriptstyle #1}{\scriptstyle #2}}}}
\def\fft#1#2{\frac{#1}{#2}}

\def\sst#1{{\scriptscriptstyle #1}}

\def\td{\tilde}

\def\dalemb#1#2{{\vbox{\hrule height .#2pt
        \hbox{\vrule width.#2pt height#1pt \kern#1pt
                \vrule width.#2pt}
        \hrule height.#2pt}}}

\newcommand{\bea}{\begin{eqnarray}}
\newcommand{\eea}{\end{eqnarray}}

\def\0{{\sst{(0)}}}
\def\1{{\sst{(1)}}}
\def\2{{\sst{(2)}}}
\def\3{{\sst{(3)}}}
\def\4{{\sst{(4)}}}
\def\5{{\sst{(5)}}}
\def\6{{\sst{(6)}}}
\def\7{{\sst{(7)}}}
\def\8{{\sst{(8)}}}

\def\R{\rlap{\rm I}\mkern3mu{\rm R}}

\def\R{\rlap{\rm I}\mkern3mu{\rm R}}

\def\R{{{\mathbb R}}}

\begin{document}

\begin{center}\ \\ \vspace{60pt}
{\Large {\bf Strings on AdS Wormholes}}\\ 
\vspace{30pt}
Mir Ali$^1$, Frenny Ruiz$^1$, Carlos Saint-Victor$^1$ and Justin F. V\'azquez-Poritz$^{1,2}$
\vspace{20pt}

{\it $^1$Physics Department\\ New York City College of Technology, The City University of New York\\ 300 Jay Street, Brooklyn NY 11201, USA.}

\vspace{10pt}
{\it $^2$The Graduate School and University Center, The City University of New York\\ 365 Fifth Avenue, New York NY 10016, USA}\\

\vspace{20pt}

{\tt am0345@bcmail.brooklyn.cuny.edu}\\
{\tt fruiz@campus.citytech.cuny.edu}\\
{\tt csaintvictor@campus.citytech.cuny.edu}\\
{\tt jvazquez-poritz@citytech.cuny.edu}

\end{center}

\vspace{30pt}

\centerline{\bf Abstract}

\noindent We consider the behavior of open strings on AdS wormholes in Gauss-Bonnet theory, which are the Gauss-Bonnet gravity duals of a pair of field theories. A string with both endpoints on the same side of the wormhole describes two charges within the same field theory, which exhibit Coulomb interaction for small separation. On the other hand, a string extending through the wormhole describes two charges which live in different field theories, and they exhibit a spring-like confining potential. A transition occurs when there is a pair of charges present within each field theory: for small separation each pair of charges exhibits Coulomb interaction, while for large separation the charges in the different field theories pair up and exhibit confinement. Two steadily-moving charges in different field theories can occupy the same location provided that their speed is less than a critical speed, which also plays the role of a subluminal speed limit. However, for some wormhole backgrounds, charges moving at the critical speed cannot occupy the same location and energy is transferred from the leading charge to the lagging one. We also show that strings on AdS wormholes in supergravity theories without higher-derivative curvature terms can exhibit these properties as well.

\thispagestyle{empty}

\pagebreak



\section{Introduction and summary}

The AdS/CFT correspondence \cite{ads1,ads2,ads3} can be used to study certain strongly-coupled gauge theories. In particular, the behavior of open strings can be related to that of particles in the field theory. The original correspondence was between type IIB string theory on AdS$_5\times S^5$ and four-dimensional maximally supersymmetric ${\cal N}=4$ $SU(N)$ Yang-Mills theory \cite{ads1}. In the large $N$ and large 't Hooft coupling $\lambda$ limit, the string theory can be approximated by classical supergravity. A large number of generalizations to the original AdS/CFT correspondence have been made \cite{ads4}. For instance, string theory on an asymptotically AdS background corresponds to a field theory undergoing a Renormalization Group flow from a conformal fixed point in the ultraviolet regime. 

A quantum theory of gravity such as string theory generally contains higher derivative corrections from stringy or quantum effects, which correspond to $1/\lambda$ or $1/N$ corrections in the field theory. However, not much is known about the precise forms of the higher derivative corrections, other than for a few maximally supersymmetric cases. The most general theory of gravity in higher dimensions that leads to second-order field equations for the metric is described by the Lovelock action \cite{lovelock}. The simplest such higher derivative theory of gravity is known as Einstein-Gauss-Bonnet theory, which only contains terms up to quadratic order in the curvature. 

Our primary motivation for considering Einstein-Gauss-Bonnet theory is that it contains Lorentzian-signature five-dimensional asymptotically locally AdS wormhole solutions which do not violate the weak energy condition, so long as the Gauss-Bonnet coupling constant is negative and bounded according to the shape of the solution \cite{gbwormholes,maeda}. In particular, we consider the static wormhole solutions that were found in \cite{troncoso1,troncoso2} which connect two asymptotically AdS spacetimes each with a geometry at the boundary that is locally $\R\times H_3$ or $\R\times S^1\times H_2$, where $H_2$ and $H_3$ are two and three-dimensional (quotiented) hyperbolic spaces, respectively. These backgrounds do not contain horizons anywhere, and the two asymptotic regions are causally connected. The proof that the disconnected boundaries must be separated by black hole horizons assumes that the Einstein equations hold \cite{theorem}, and therefore does not automatically carry over to higher-derivative gravity theories such as Einstein-Gauss-Bonnet theory.

If one is able to apply the standard AdS/CFT correspondence to these AdS wormholes, then string theory on such a background corresponds to two quantum field theories (assuming that each boundary admits a well-defined field theory). Each theory undergoes a Renormalization Group flow from a conformal fixed point in its UV limit. However, we would like to emphasize that the AdS/CFT correspondence has not actually been tested for gravitational backgrounds of Einstein-Gauss-Bonnet theory and may not be valid. We are taking some liberty in applying the standard AdS/CFT correspondence in this context, since the duality is between a gauge theory and a ten-dimensional string theory, whereas these five-dimensional backgrounds have not been embedded within string theory. We are making the working assumption that the low-energy effective five-dimensional description of gauge theory/string theory duality has a sensible derivative expansion in which the higher curvature terms are systematically suppressed by powers of the Planck length. After making the appropriate field redefinitions, the curvature-squared terms in the action appear in the Gauss-Bonnet combination. Thus, within the limitations of the derivative expansion, one can use the AdS/CFT correspondence to determine the properties of gauge theories dual to Einstein-Gauss-Bonnet gravitational backgrounds \cite{buchel}.

In fact, one might expect that within the vast string landscape there are higher derivative corrections which lead to similar backgrounds with asymptotically AdS regions with multiple disconnected boundaries. Concrete examples of such backgrounds already exist within string theory, including a multi-boundary orbifold of AdS$_3$ \cite{balasubramanian} and the Maoz-Maldacena wormhole \cite{maoz}. Although the orbifold of AdS$_3$ clearly has lower dimensionality while the Maoz-Maldacena wormhole has Euclidean signature, these backgrounds serve as toy models with which to demonstrate that open strings apparently exhibit some fairly universal behavior on AdS wormhole type backgrounds. This indicates that our findings regarding the behavior of open strings on AdS wormholes in Einstein-Gauss-Bonnet theory could also apply to similar backgrounds which may arise in string theory.

In this paper, we consider the behavior of open strings on these AdS wormholes. The string endpoints lie on probe D-branes and correspond to charges within one or the other of the field theories. We refer to these charges as type 1 and type 2 charges. If both of the string endpoints lie on the same D-brane and are both located on the same side of the wormhole, then this corresponds to a type 1 or a type 2 charge-anticharge pair. On the other hand, if a string is extended through the neck of the wormhole such that its endpoints lie on opposite sides, then this corresponds to one of the charges being type 1 while the other is type 2.

Technically, we should also consider the quantum fluctuations of the string worldsheet \cite{drukker}, which formally dominate over the higher derivative corrections of the background. However, we have checked that the corrections to the string worldsheet do not change the qualitative behavior of strings, which is the focus of our interest.

\paragraph{Summary.}

The expectation value of a rectangular Wilson loop can be computed from the proper area of the string worldsheet \cite{wilson1,wilson2}, from which the energy of a pair of charges can be read off. Applying this prescription to AdS wormholes, one can recover the result that a pair of charges of the same type exhibit Coulombic attraction in the UV limit of the field theory being probed by a string which is far from the neck of the wormhole. We generalize this prescription for the case of two charges which do not live within the same field theory. Even though the string endpoints lie on opposite sides of the wormhole, there is a rectangular contour within the physical spacetime whose horizontal sides run along the distance between the two charges and whose vertical sides run along time. We find that two charges of different type exhibit confinement with a spring-like potential. Thus, the strength of the interaction can be parameterized in terms of an effective force constant. A pair of heavy charges exhibits a weaker interaction than do lighter charges. We also find that these characteristics are shared for open strings on a multi-boundary orbifold of AdS$_3$ \cite{balasubramanian} as well as on the Maoz-Maldacena wormhole \cite{maoz}. For a family of AdS wormholes in Einstein-Gauss-Bonnet theory \cite{troncoso1,troncoso2}, the effective force constant can be tuned by adjusting a parameter associated with the apparent mass on each side of the wormhole.

We find that there is a rather curious transition which involves a quadruplet of charges that consists of a charge-anticharge pair within each field theory. For small charge-anticharge separations, each pair exhibits Coulomb interaction and does not interact with the pair of charges in the other field theory. However, for large separation, it is the charges of different types that interact and exhibit confinement whereas the pairs of charges of the same type no longer interact with each other. In this sense, each pair of charges of the same type exhibits the feature of effectively having a screening length, even though both field theories are at zero temperature. 

We also consider the behavior of steadily-moving strings. In particular, demanding that the string configuration is timelike imposes constraints on the mass parameters of the corresponding charges, as well as on their separation. We find that the charges have an upper bound on their speed which depends on the mass parameter of one of the charges (which charge depends on a parameter associated with the wormhole geometry). This speed limit is generally less than the speed of light, which is a result of the fact that the proper velocity of the string endpoints is greater than the physical velocity in the field theory \cite{argyres}.

In addition, we consider the behavior of strings on an Einstein-Gauss-Bonnet wormhole which connects an asymptotically locally AdS spacetime with another nontrivial smooth spacetime at the other asymptotic region. Even though one of the asymptotic regions of the Einstein-Gauss-Bonnet wormhole is not AdS, we find that open strings on this background share many of the same characteristics as open strings on a multi-boundary orbifold of AdS$_3$ \cite{balasubramanian}. This may be an indication that some aspects of the behavior of open strings on wormholes are fairly universal, regardless of the asymptotic geometries. 

Namely, we find that a pair of steadily-moving charges of different types can occupy the same location with no transfer of energy or momentum between them, provided that their speed is less than a certain critical speed. This critical speed coincides with the speed limit of the charges and monotonically decreases with the mass parameters. If the charges move at this critical speed, then they exhibit a separation gap which increases with the mass parameter of the lagging charge but is not so sensitive to the mass parameter of the leading charge. Also at this critical speed, energy and momentum are transferred from the leading charge to the lagging one. Note that the procedure for calculating the rate of energy and momentum transfer parallels the way in which the rate of energy loss of a quark moving through a strongly-coupled plasma was calculated via a dragging string moving on a five-dimensional AdS black hole background \cite{herzog,teaney,gubser}. Curvature-squared corrections of the background were considered in \cite{squared1,squared2,squared3}. However, the case of an AdS black hole corresponds to energy being transferred from the quark to the plasma, whereas in the present case of an AdS wormhole the energy is transferred between two different types of charges. We find that the rate at which energy is transferred actually {\it decreases} with the mass parameter of the lagging charge. While this is not necessarily an unexpected feature of a wormhole background, it is a rather surprising characteristic on the field theory side. 

Wormholes of various dimensionality which connect AdS$_n\times S^m$ in one asymptotic region to flat spacetime in the other have been constructed in Einstein gravity \cite{lu1,lu2}. It would be interesting to see if open strings on these wormholes have similar features to what we have found in this paper, even though the notion of time tends to be different for the two asymptotic regions. 

This paper is organized as follows. In section 2, we provide general formulae for the AdS wormhole metric, string embeddings and equations of motion. In section 3, we analyze the behavior of static string configurations on a family of AdS wormholes that arise in Einstein-Gauss-Bonnet theory. In section 4, we consider steadily-moving string configurations on these AdS wormhole backgrounds. In section 5, we consider strings on an Einstein-Gauss-Bonnet wormhole for which only one asymptotic region is AdS.
Lastly, in section 6, we consider strings on AdS wormholes that can be embedded within string theory.

\section{Generalities}

\subsection{Wormholes in five-dimensional Einstein-Gauss-Bonnet theory}

We are making the assumption that the low-energy effective five-dimensional description of gauge theory/string theory duality has a sensible derivative expansion \cite{buchel}. After field redefinitions, the leading terms in the action take on the Einstein-Gauss-Bonnet form:
\be\label{gbaction}
I=\kappa\int d^5x \sqrt{-g} \left[ \fft{12}{\ell^2}+R+\alpha\ell^2 (R^2-4R_{\mu\nu} R^{\mu\nu}+R_{\alpha\beta\gamma\delta} R^{\alpha\beta\gamma\delta})+\cdots\right],
\ee
where $\kappa$ is related to the Newton constant, $\ell$ corresponds to the AdS curvature scale at leading order, and $\alpha$ is the dimensionless Gauss-Bonnet coupling. In order to make the above derivative expansion, we require that $\alpha\ll 1$. For a fixed value of $\alpha$, there is a family of wormhole solutions to (\ref{gbaction}) which is described by the metric \cite{troncoso1,troncoso2}
\be\label{metric1}
ds^2=\ell^2 \Big(-\cosh^2(\rho-\rho_0) dt^2+d\rho^2+\cosh^2\rho\ d\Sigma_3^2\Big)\,.
\ee
The base manifold $\Sigma_3$ can be either $H_3/\Gamma$ or $S^1\times H_2/\Gamma$, where $H_3$ and $H_2$ are hyperbolic spaces and $\Gamma$ is a freely acting discrete subgroup of $O(3,1)$ and $O(2,1)$, respectively. The hyperbolic part of the base manifold must be quotiented so that this geometry describes a wormhole rather than a gravitational soliton with a single conformal boundary.
The neck of the wormhole is located at $\rho=0$, which connects two asymptotically locally AdS regions at $\rho\rightarrow\pm\infty$. The geometry is devoid of horizons and radial null geodesics connect the two asymptotic regions in a finite time $\Delta t=\pi$. These wormholes can evade the proof that the disconnected boundaries must be separated by black hole horizons \cite{theorem}, since we are dealing with a higher-derivative theory of gravity. No energy conditions are violated by these wormholes, since the stress-energy tensor vanishes everywhere \cite{troncoso1,troncoso2}. The stability of these wormholes against scalar field perturbations has been discussed in \cite{stability}.

It has been shown that gravity pulls towards a fixed hypersurface at $\rho=\rho_0$, which lies in parallel with the neck of the wormhole \cite{troncoso1,troncoso2}. In particular, timelike geodesics are confined, and oscillate about this hypersurface. Although the wormhole is massless, for nonzero $\rho_0$, the mass of the wormhole appears to be positive for observers located at on one side and negative for the other. The parameter $\rho_0$ is related to the apparent mass on each side of the wormhole; for $\rho_0=0$, the wormhole exhibits reflection symmetry. 

If we can apply the AdS/CFT correspondence to this background, then this is the gravity dual of two interacting field theories on $\R\times \Sigma_3$. Except in the UV limit, the conformal symmetry of both theories is broken by the length scale associated with $\Sigma_3$ ($H_3$ and $H_3$ must have the radius $1$ and $1/\sqrt{3}$, respectively) as well as by the parameter $\rho_0$. By computing the boundary stress tensors \cite{stresstensor}, one can determine how $\rho_0$ affects the expectation value of the stress tensor of each field theory.

\subsection{String embeddings and equations of motion}

The dynamics of a classical string on a spacetime with the metric $G^{\mu\nu}$ are governed by the Nambu-Goto action
\be
S=-T_0 \int d\sigma d\tau \sqrt{-g}\,,
\ee
where $(\tau,\sigma)$ are the worldsheet coordinates, $g_{ab}$ is the induced metric and $g=\mbox{det} (g_{ab})$. For a map $X^{\mu}(\tau,\sigma)$ from the worldsheet into spacetime,
\be
-g=(\dot X\cdot X^{\prime})^2-(X^{\prime})^2 (\dot X)^2\,.
\ee
where $\dot X\equiv\partial_{\tau}X$, $X^{\prime}\equiv\partial_{\sigma}X$ and $\dot X\cdot X^{\prime}=\dot X^{\mu} (X^{\nu})^{\prime} G_{\mu\nu}$. We will consider a string which is localized at a point in $H_2$ and moves along the $S^1$ direction, which we label $x$. Choosing a static gauge where $\sigma=\rho$ and $\tau=t$, we find that the equation of motion for a string on the background metric (\ref{metric1}) is given by
\be\label{eom}
\fft{\partial}{\partial\rho} \Big( \fft{\cosh^2(\rho-\rho_0) \cosh^2\rho\ x^{\prime}}{\sqrt{-g}}\Big)-\cosh^2\rho\ \fft{\partial}{\partial t} \Big( \fft{\dot x}{\sqrt{-g}}\Big)=0\,,
\ee
where
\be\label{g}
-\fft{g}{\ell^4}=\cosh^2(\rho-\rho_0)+\cosh^2(\rho-\rho_0) \cosh^2\rho\ x^{\prime 2}-\cosh^2\rho\ \dot x^2\,.
\ee

It will also be useful to express the equation of motion in the alternative gauge $\sigma=x$ and $\tau=t$, for which
\bea\label{eomalt}
\fft{\partial}{\partial x}\left( \fft{\cosh^2 (\rho-\rho_0) \rho^{\prime}}{\sqrt{-g}}\right) &-& \cosh^2\rho \fft{\partial}{\partial t} \left( \fft{\dot\rho}{\sqrt{-g}}\right) = \cosh\rho\cosh (\rho-\rho_0)\sinh (2\rho-\rho_0)\nn\\
&-& \cosh\rho\sinh\rho\ \dot\rho^2+\cosh (\rho-\rho_0)\sinh (\rho-\rho_0) \rho^{\prime 2}\,,
\eea
where
\be\label{galt}
-\fft{g}{\ell^4}=\cosh^2\rho\cosh^2 (\rho-\rho_0)-\cosh^2\rho\ \dot\rho^2+\cosh^2 (\rho-\rho_0)\rho^{\prime 2}\,.
\ee

The general expressions for the canonical momentum densities of the string are given by
\bea\label{currents}
\pi_{\mu}^0 &=& -T_0\ G_{\mu\nu} \fft{(\dot X\cdot X^{\prime})(X^{\nu})^{\prime}-(X^{\prime})^2(\dot X^{\nu})}{\sqrt{-g}}\,,\nn\\
\pi_{\mu}^1 &=& -T_0\ G_{\mu\nu} \fft{(\dot X\cdot X^{\prime})(\dot X^{\nu})-(\dot X)^2(X^{\nu})^{\prime}}{\sqrt{-g}}\,.
\eea
The energy density and density of the $x$-component of momentum on the string worldsheet are given by $\pi_t^0$ and $\pi_x^0$, respectively. Thus, the total energy and momentum of the string are
\be\label{generalEp}
E=-\int d\sigma\ \pi_t^0\,,\qquad p=\int d\sigma\ \pi_x^0\,.
\ee
We will also make use of the fact that the string tension is given by $T_0=\sqrt{\lambda}/(2\pi \ell^2)$, where $\lambda$ is the 't Hooft coupling of the field theory.

\section{Static strings}

\subsection{Straight strings}

The simplest solution is a constant $x=x_0$, for which the string is straight, as shown in Figure 1. Suppose that the endpoints are located at $\rho=-\rho_1$ and $\rho=\rho_2$, where we will always take $\rho_1, \rho_2>0$. We will refer to the charges associated with the endpoints at $\rho=-\rho_1$ and $\rho=\rho_2$ as a type 1 charge (blue) and a type 2 charge (red), respectively. Also, we will refer to the  radial locations of the string endpoints $\rho_1$ and $\rho_2$ as the mass parameters of the charges.
The D-branes (or at least the part that lies closest to the neck of the wormhole) are denoted by the aqua-colored surfaces and the grey surface represents the fixed hypersurface at $\rho=\rho_0$.
\begin{figure}[ht]
   \epsfxsize=2.7in \centerline{\epsffile{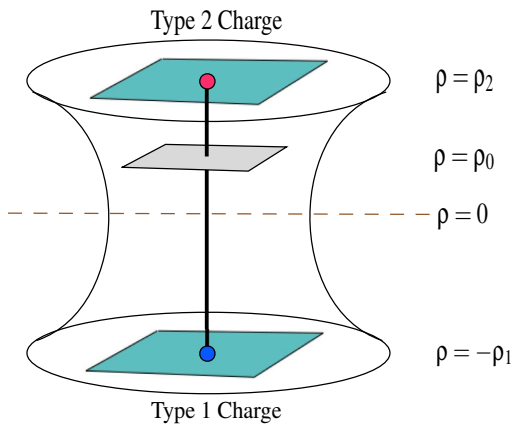}}
   \caption[FIG. \arabic{figure}.]{\footnotesize{A static string stretching straight through a wormhole.}}
\label{fig1}
\end{figure}
From (\ref{generalEp}), we find that the static string has vanishing momentum, and its energy is given by
\be\label{restE}
E=T_0 \ell^2\big( \sinh(\rho_1+\rho_0)+\sinh (\rho_2-\rho_0)\big)\,.
\ee

From the equation of motion (\ref{eomalt}) expressed in the alternative gauge, we find that there is also a solution at constant $\rho=\rho_0/2$. However, since this would presumably correspond to an extended object in the field theory, we will not consider this string configuration further.

\subsection{Curved strings}

\begin{figure}[ht]
\begin{center}
$\begin{array}{c@{\hspace{1.25in}}c}
\epsfxsize=1.8in \epsffile{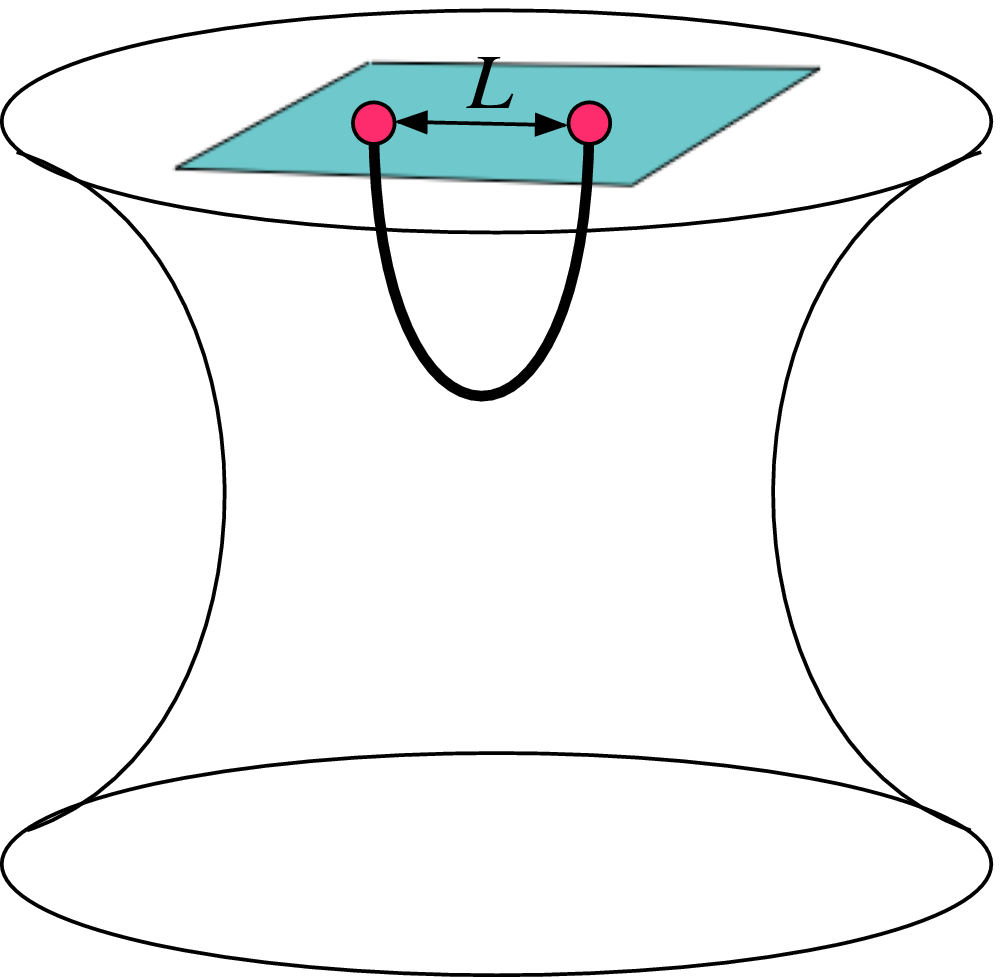} &
\epsfxsize=1.8in \epsffile{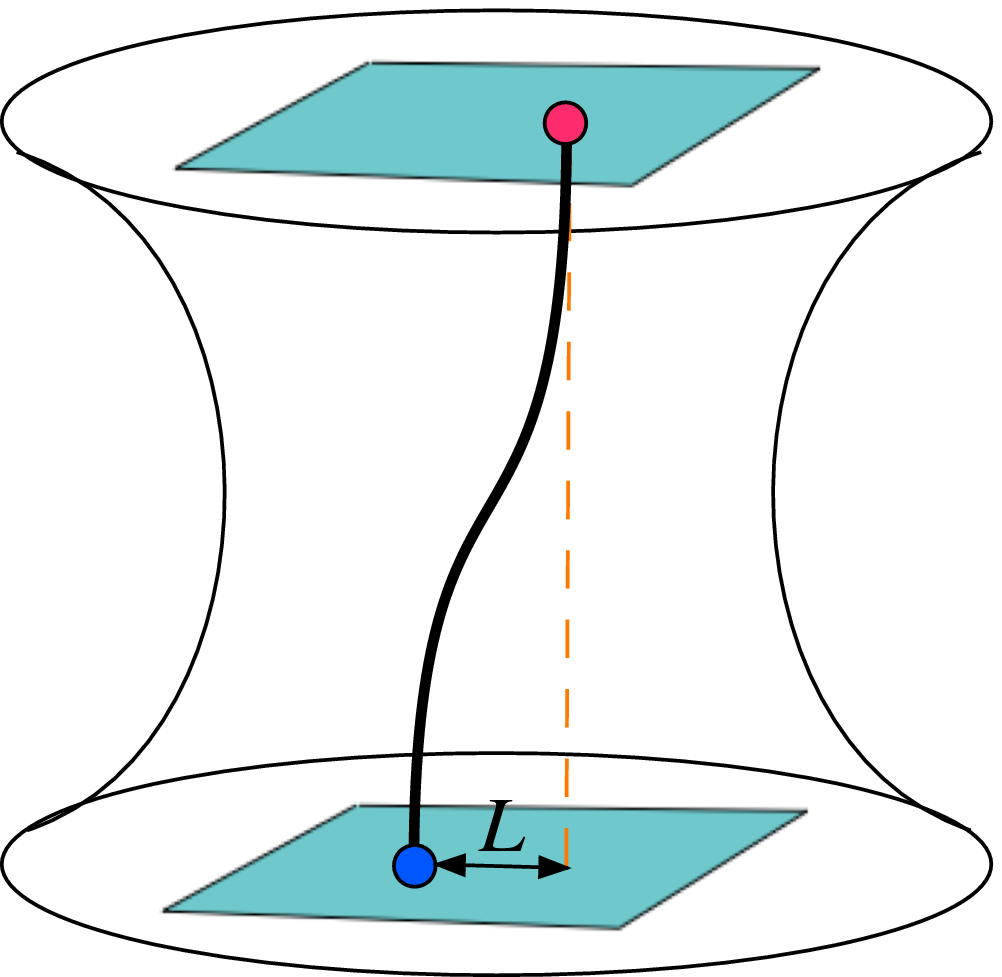}
\end{array}$
\end{center}
\caption[FIG. \arabic{figure}.]{\footnotesize{A string with both endpoints on the same side of the wormhole (left) and on opposite sides (right).}}
   \label{fig4}
\end{figure}

Curved strings can either have both endpoints on the same side of the wormhole or else on opposite sides, as shown in Figure \ref{fig4}. 
We will first consider the first case, which corresponds to a pair of charges of the same type. This means that the string dips down towards the wormhole and then comes back up. We will denote the turning point of the string by $\rho_t$, and $L$ is the distance between the endpoints of the string. For simplicity, we will take the endpoints to be at $\rho=\rho_2$, so that we have a pair of type 2 charges. For $\rho_2$, $\rho_t\gg 1$, the string is not sensitive to the presence of the wormhole and so it can be found that its energy goes as $1/L$. A Coulomb potential energy is consistent with the fact that the theory is conformal in the extreme UV regime. If the masses associated with the charges are decreased, or if the distance between the charges is increased, then the string dips closer to the wormhole. Then terms in the energy that are of higher order in $L$ will become important.

From (\ref{eom}) and (\ref{g}), we find the distance $L$ between the string endpoints is given by
\be
L=2\cosh (\rho_t-\rho_0) \cosh\rho_t \int_{\rho_t}^{\rho_2} \fft{d\rho}{\cosh\rho \sqrt{\cosh^2 (\rho-\rho_0) \cosh^2\rho-\cosh^2 (\rho_t-\rho_0)\cosh^2\rho_t}}\,.
\ee
The above integration region assumes that $\rho_2>\rho_t$. As we will see shortly, there are string solutions for which $\rho_2<\rho_t$, in which case the upper and lower integration bounds must be interchanged.

\begin{figure}[ht]
\begin{center}
$\begin{array}{c@{\hspace{1.0in}}c}
\epsfxsize=2.6in \epsffile{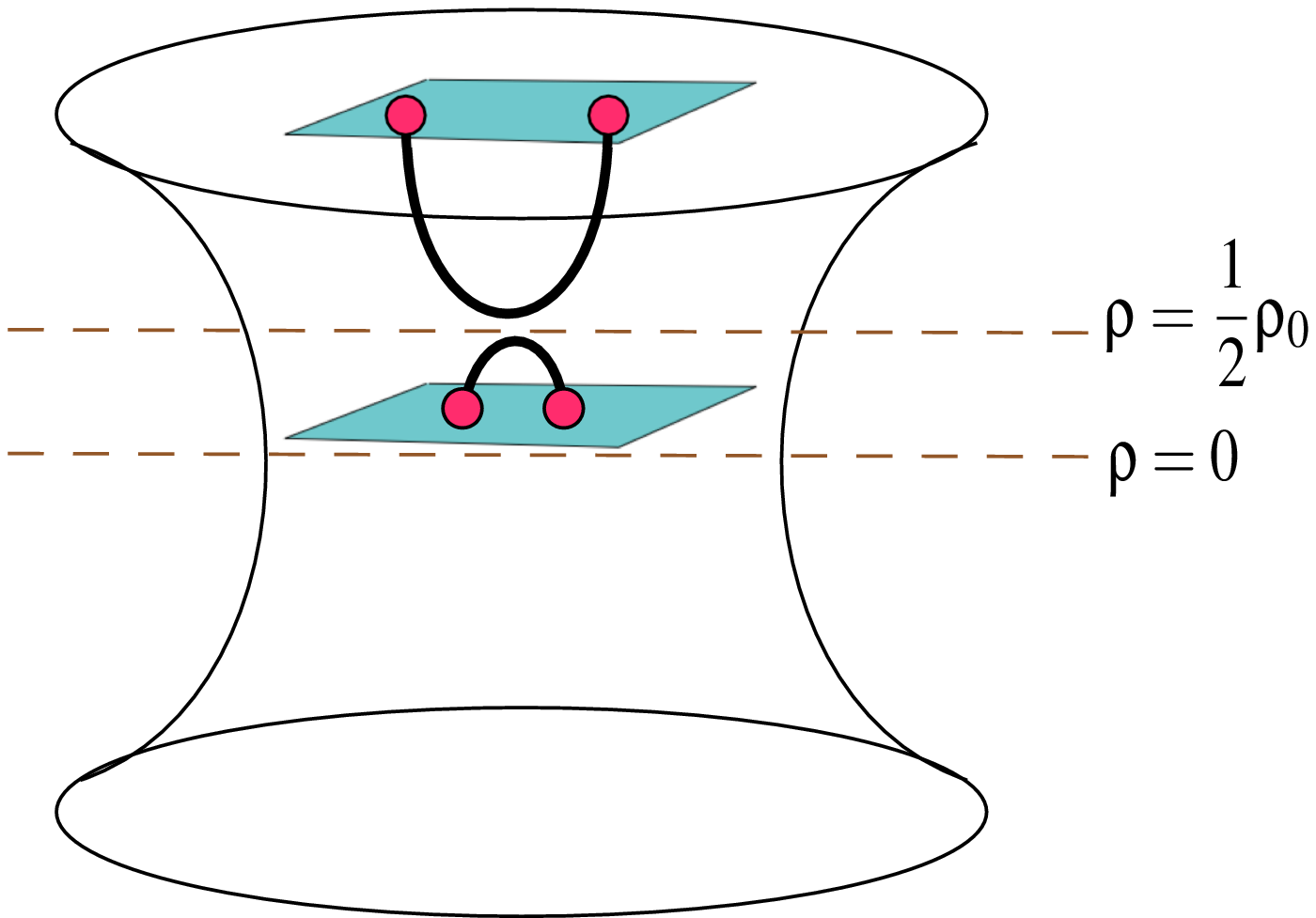} &
\epsfxsize=2.6in \epsffile{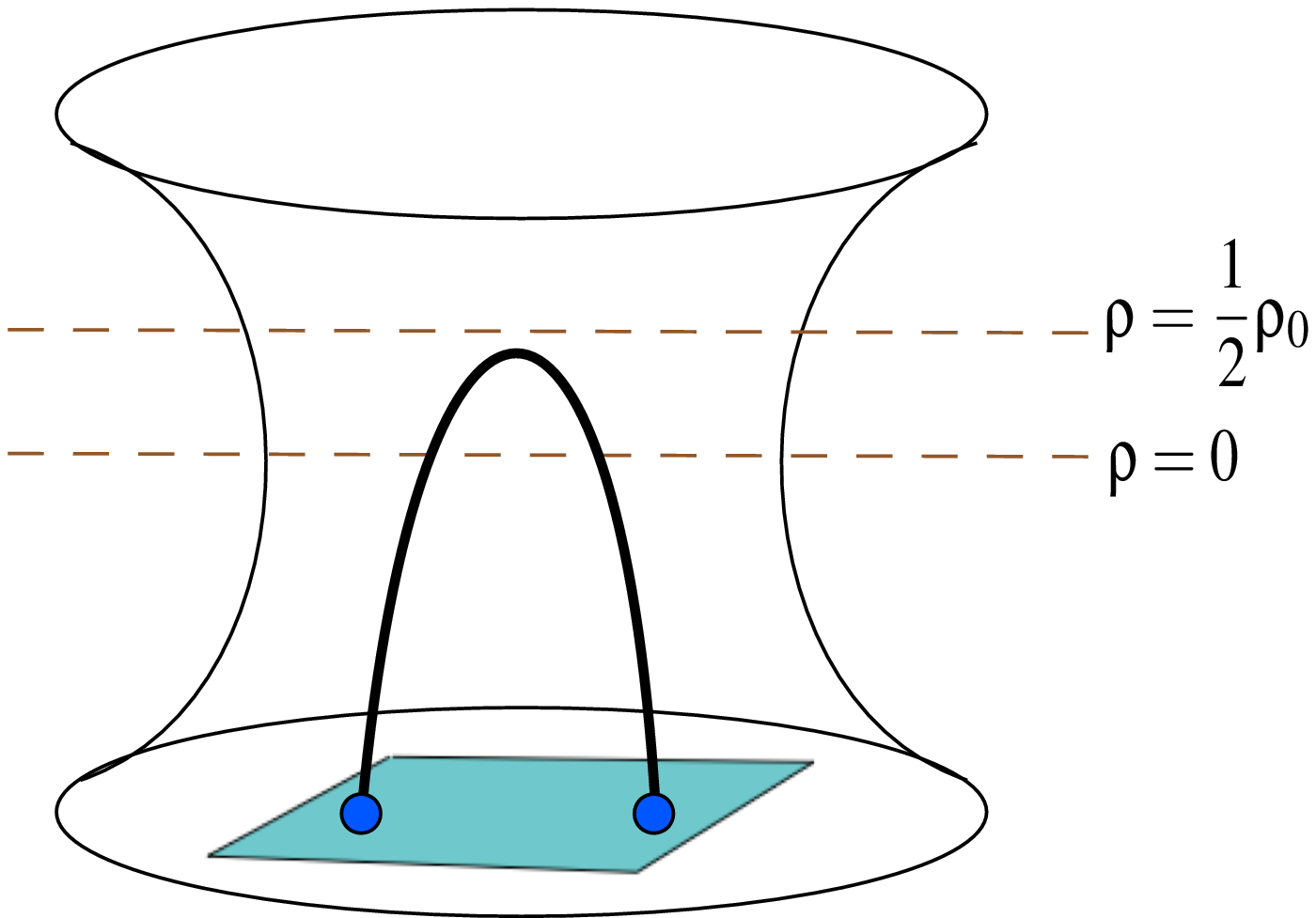}
\end{array}$
\end{center}
\caption[FIG. \arabic{figure}.]{\footnotesize{Strings bend towards $\rho=\ft12\rho_0$ regardless of where the string endpoints are located, even if this means that the string bends away from the neck of the wormhole (left) or goes through the neck of the wormhole (right).}}
   \label{otherside}
\end{figure}

For $\rho_2>\ft12 \rho_0$, the reality of $L$ implies that the turning point is located in the interval $\rho_2>\rho_t\ge\ft12 \rho_0$. Similarly, a string with endpoints at $\rho_2<\ft12\rho_0$ has a turning point in the interval $\rho_2<\rho_t\le\ft12 \rho_0$. This means that if the turning points are within the interval $0<\rho_2<\rho_0/2$, then the string will bend {\it away} from the center of the wormhole. Moreover, if the endpoints are located on the opposite side of the wormhole as the hypersurface $\rho=\rho_0$, then there are strings which go through the neck of the wormhole and have a midsection on the opposite side of the wormhole as its endpoints. All of these possibilities are shown in Figure \ref{otherside}. 

\begin{figure}[ht]
   \epsfxsize=3.0in \centerline{\epsffile{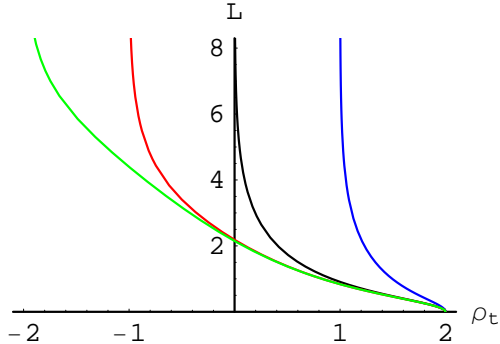}}
   \caption[FIG. \arabic{figure}.]{\footnotesize{The distance between string endpoints $L$ versus the turning point $\rho_t$ for the case in which both endpoints are on the same side of the wormhole. We have set $\rho_2=2$ and $\rho_0=2$ (blue), $0$ (black), $-2$ (red) and $-4$ (green).}}
\label{fig3}
\end{figure}

As shown in Figure \ref{fig3} for various values of $\rho_0$, the turning point $\rho_t$ monotonically decreases with $L$. In particular, as $L$ gets large the turning point asymptotically approaches $\ft12 \rho_0$ from above. Note that, unlike the case of strings on an AdS black hole background, there is a single string configuration for each value of $L$. 
From the field theory perspective, this means that the charges do not exhibit a screening length but rather remain interacting no matter how far they are from each other (which is not the case in the presence of additional pairs of charges, as we will discuss shortly). In particular, as the distance between the charges increases, the inter-charge potential becomes ever more sensitive to the IR characteristics of that sector of the theory. For vanishing $\rho_0$ (black curve in Figure \ref{fig3}), the potential is partly determined by the extreme IR region of the theory for large $L$. However, for $\rho_0>0$, the large-distance potential only reflects the features of the energy range corresponding to the region $\rho\ge \ft12 \rho_0$ (red curve in Figure \ref{fig3} is for $\rho_0=2$). 

We will now consider a curved string with endpoints on opposite sides of the wormhole, which corresponds to a type 1 charge and a type 2 charge. From (\ref{eom}) and (\ref{g}), the distance between the string endpoints in the $x$ direction is
\be
L=C \int_{-\rho_1}^{\rho_2} \fft{d\rho}{\cosh\rho\sqrt{\cosh^2(\rho-\rho_0)\cosh^2\rho-C^2}}\,.
\ee
For small separation $L$, we can expand in $C\ll 1$ so that
\be\label{LC}
L\approx C \int_{-\rho_1}^{\rho_2} \fft{d\rho}{\cosh (\rho-\rho_0) \cosh^2\rho}\,.
\ee

\begin{figure}[ht]
\begin{center}
$\begin{array}{c@{\hspace{0.6in}}c}
\epsfxsize=2.8in \epsffile{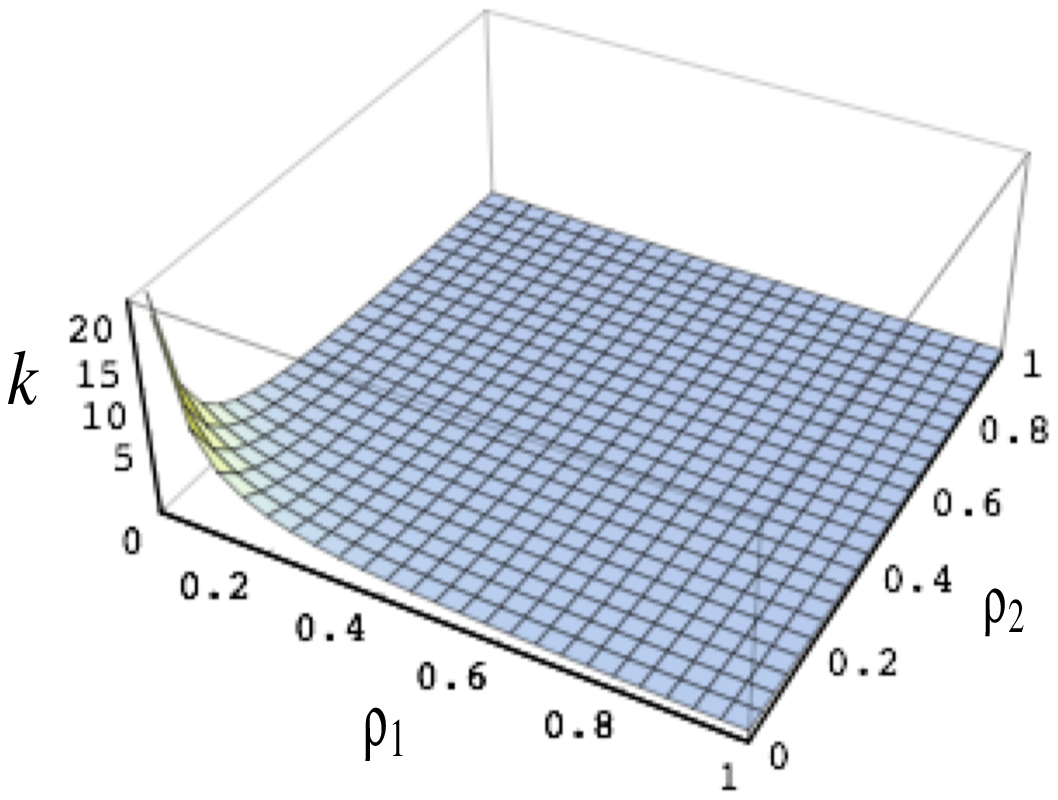} &
\epsfxsize=2.8in \epsffile{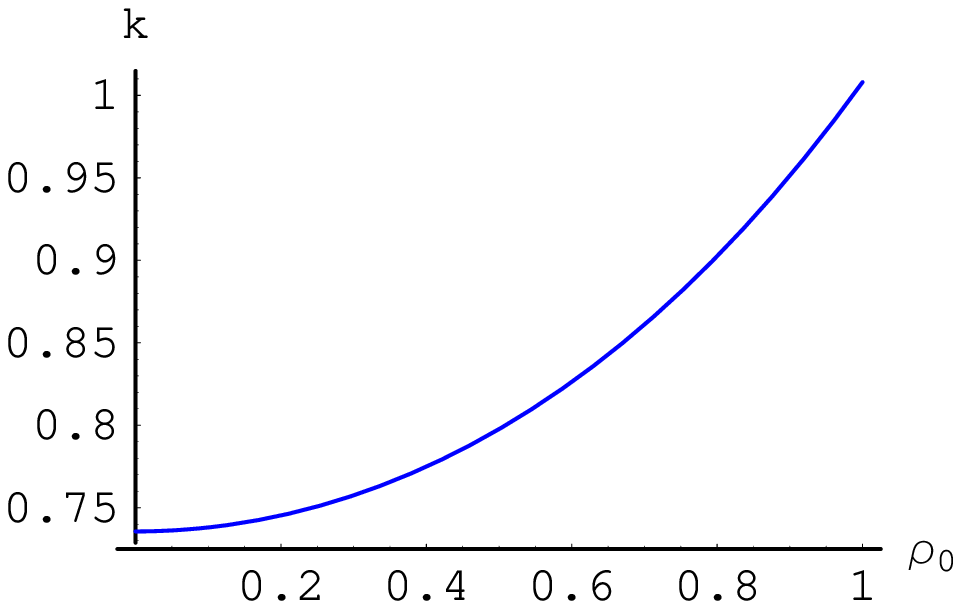}
\end{array}$
\end{center}
\caption[FIG. \arabic{figure}.]{\footnotesize{The left plot shows the effective force constant $k$ versus $\rho_1$ and $\rho_2$ for $\rho_0=0.1$. The right plot shows $k$ versus $\rho_0$ for $\rho_1=\rho_2=1$.}}
   \label{fig12}
\end{figure}

We will now Euclideanize the metric (\ref{metric1}) in order to compute the energy of the string configuration in terms of its action. The action for the string Euclidean worldsheet is given by
\be\label{EuclideanAction}
S_E=T_0\int d\sigma d\tau \sqrt{g}\,.
\ee
The energy of a pair of charges is $E=S_E/\Delta t$, where we take the time interval $\Delta t\rightarrow\infty$. Then for a static string (\ref{eom}) and (\ref{g}) become
\be
x^{\prime}=\fft{C\sqrt{g}}{\cosh^2(\rho-\rho_0)\cosh^2\rho}\,,\qquad \fft{g}{\ell^4}=\cosh^2(\rho-\rho_0)\Big( 1+\cosh^2\rho\ x^{\prime 2}\Big)\,.
\ee
We find the energy of the string to be
\be
E=T_0\ell^2 \int_{-\rho_1}^{\rho_2} d\rho\ \fft{\cosh^2(\rho-\rho_0)\cosh\rho}{\sqrt{\cosh^2(\rho-\rho_0)\cosh^2\rho-C^2}}\,.
\ee
In the small separation limit $C\ll 1$,
\be
E=E_{straight}+\fft12 T_0\ell^2 C^2 \int_{-\rho_1}^{\rho_2} \fft{d\rho}{\cosh (\rho-\rho_0)\cosh^2\rho}\,,
\ee
where $E_{straight}$ is the energy of a straight string given by (\ref{restE}). Using (\ref{LC}) to express $C$ in terms of $L$, we find
\be
E=E_{straight}+\fft12 kL^2\,,
\ee
where
\be
k=\fft{\sqrt{\lambda}}{2\pi} \Big( \int_{-\rho_1}^{\rho_2} \fft{d\rho}{\cosh (\rho-\rho_0)\cosh^2\rho}\Big)^{-1}\,,
\ee
and we have used $T_0=\sqrt{\lambda}/(2\pi\ell^2)$. Thus, for small $L$, a pair of charges of opposite types are confined and exhibit a spring-like potential, as opposed to a pair of charges of the same type which exhibit a Coulomb potential. Since we are at zero temperature, there is no screening length. The left plot in Figure \ref{fig12} shows that the effective force constant $k$ monotonically decreases with the mass parameters $\rho_1$ and $\rho_2$, where we have taken $\rho_0=0.1$. This means that heavy charges are not as sensitive to the confining potential as lighter charges.
We also find that $k$ monotonically increases with the $\rho_0$ parameter, as shown in the right plot of Figure \ref{fig12} for $\rho_1=\rho_2=1$.

\begin{figure}[ht]
\begin{center}
$\begin{array}{c@{\hspace{1.25in}}c}
\epsfxsize=1.8in \epsffile{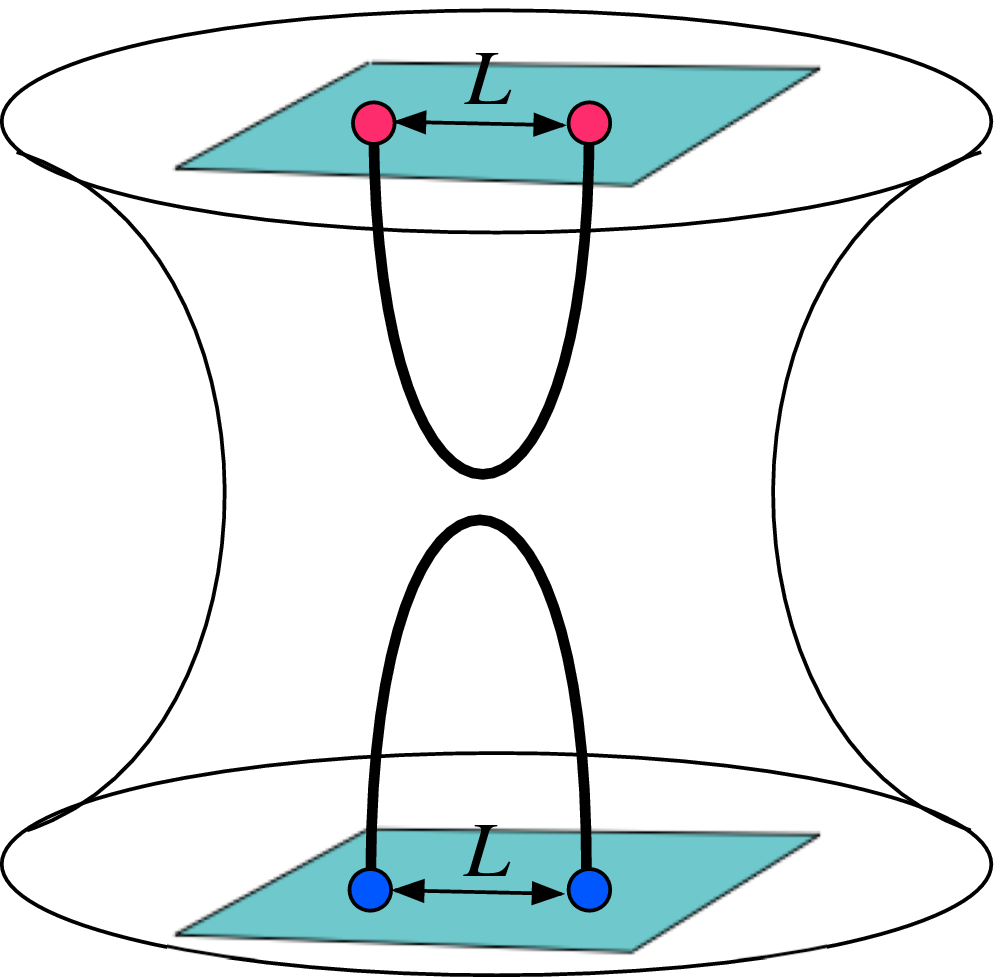} &
\epsfxsize=1.8in \epsffile{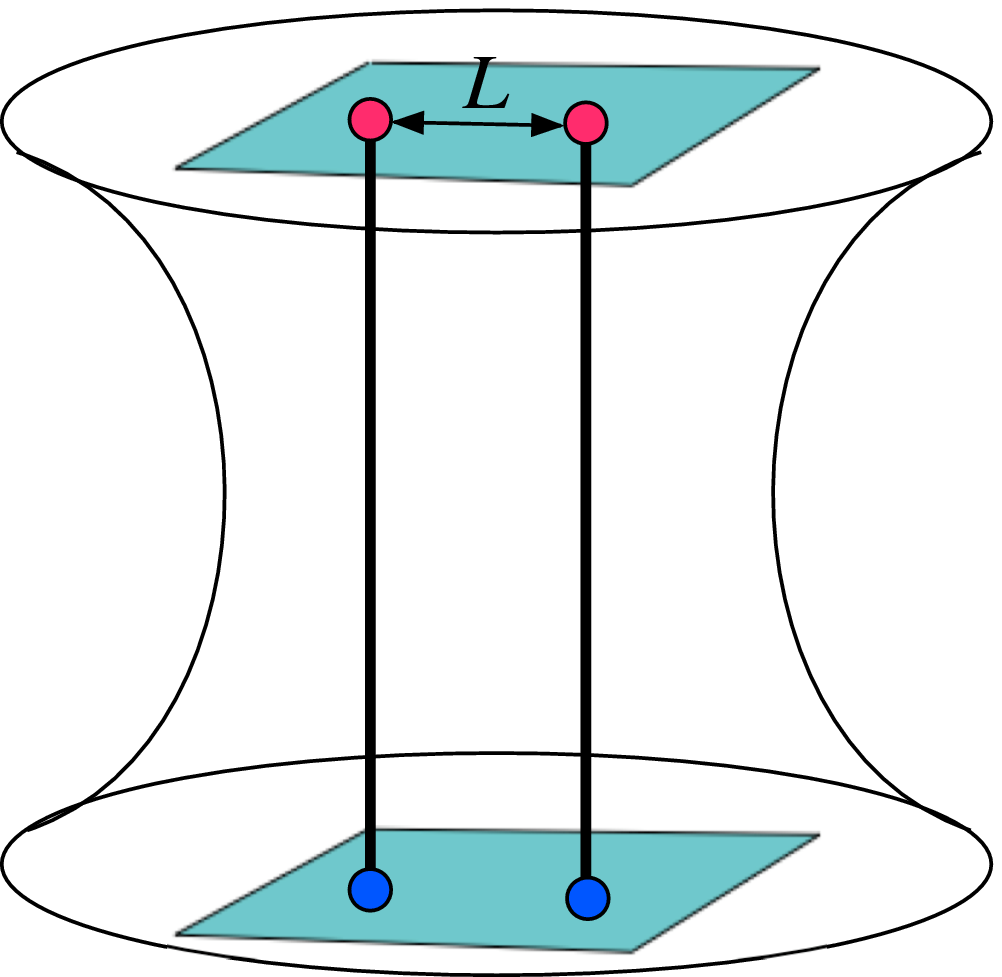}
\end{array}$
\end{center}
\caption[FIG. \arabic{figure}.]{\footnotesize{There is a critical distance $L_{crit}$ associated with two sets of charge-anticharge pairs. For $L<L_{crit}$, the left configuration is energetically favorable while, for $L>L_{crit}$, the right configuration is the favored one.}}
   \label{figtrans}
\end{figure}

For two sets of charge-anticharge pairs, an interesting transition can take place. For instance, suppose we have a pair of interacting type 1 charges and a pair of interacting type 2 charges, described by the two curved strings shown in the left plot of Figure \ref{figtrans}. For simplicity, we will assume that both pairs of charges are a distance $L$ apart. For small $L$, there is no interaction between the type 1 pair and the type 2 pair. However, as $L$ increases, the strings bend closer to each other. Note that these strings cannot pass each other, since neither one can pass the radius $\rho=\rho_0/2$. However, even before either string hits this bound, the string configuration in the right plot of Figure \ref{figtrans} becomes the energetically favorable one with the same set of endpoints. Thus, for large $L$, each type 1 charge becomes coupled to a type 2 charge. This is similar to a screening length, in that the charges of the same type no longer interact with each other. However, since we are at zero temperature, this is really a feature of having multiple pairs of charges close to each other.

In principle, there is no need for us to assume that each pair of charges are separated by the same distance $L$. Suppose that $L_1$ and $L_2$ denote the distance between the type 1 charges and the type 2 charges, respectively. Then, in general, there is a critical admixture of $L_1$ and $L_2$ for which the aforementioned transition takes place. The only difference is that, past the critical lengths, each string ``connecting" a type 1 charge to a type 2 charge is now curved, rather than straight, as shown in the right plot of Figure \ref{fig4}. If the endpoints are not held in place by an external force, then each string will straighten out.

Note that this type of transition can occur even if there are two pairs of charges that are all of the same type, such as is shown in Figure \ref{otherside}. This indicates that such processes may not be restricted only to field theories described by AdS wormholes.

\section{Steadily-moving strings}

\subsection{Straight strings}

A steadily-moving straight string is given by $x(t,\rho)=x_0+vt$. For this solution, we find that
\be
-\fft{g}{\ell^4}=\cosh^2(\rho-\rho_0)-v^2 \cosh^2\rho\,.
\ee
For simplicity, we will take $\rho_0\ge 0$ in the remainder of the paper. Then for $v< v_{crit}\equiv e^{-\rho_0}$, $-g>0$ everywhere for $v<1$, which implies that no parts of the string move faster than the local speed of light. This is to be contrasted with a straight string steadily moving in an AdS black hole background \cite{herzog,gubser} which, for any nonzero velocity, has a region with $-g<0$ and is therefore not a physical solution. On the other hand, for $v=v_{crit}$, we require that $\rho_2<\rho_0/2$. For $v> v_{crit}$, $g=0$ at $\rho=\rho_{crit}$ where
\be\label{pcrit}
\rho_{crit}=\ln \sqrt{\fft{e^{\rho_0}-v}{v-e^{-\rho_0}}}\,,
\ee
at which point the induced metric on the string worldsheet is degenerate. We can still have steadily-moving straight strings which lie entirely within the region $\rho_{crit}<\rho$. However, for strings which have a region that lies within $\rho<\rho_{crit}$, $-g<0$ and the action, energy and momentum are complex. This is a signal that this part of the string travels faster than the local speed of light and must be discarded. In the next section, we will consider curved strings with $v>v_{crit}$ which satisfy $-g>0$.

For the special case of $\rho_0=0$, we have
\be
-\fft{g}{\ell^4}=(1-v^2) \cosh^2\rho>0\,,
\ee
for all $v<1$. In other words, for this case the critical speed $v_{crit}=1$ for which $g=0$ and the induced metric on the string worldsheet is degenerate.

The energy and momentum of the steadily moving string are given by the integrals
\bea\label{Ep}
E &=& T_0\ell^2 \int_{-\rho_1}^{\rho_2} d\rho\ \fft{\cosh^2 (\rho-\rho_0)}{\sqrt{\cosh^2(\rho-\rho_0)-v^2 \cosh^2\rho}}\,,\nn\\ 
p &=& T_0\ell^2v \int_{-\rho_1}^{\rho_2} d\rho\ \fft{\cosh^2 \rho}{\sqrt{\cosh^2(\rho-\rho_0)-v^2 \cosh^2\rho}}\,.
\eea
For the special case of $\rho_0=0$, we find that 
\be
E=\fft{E_{rest}}{\sqrt{1-v^2}}\,,\qquad E_{rest}=T_0\ell^2 (\sinh\rho_1+\sinh\rho_2)\,.
\ee
For non-vanishing $\rho_0$, the expressions for $E$ and $p$ cannot be integrated in closed form. Note that, since $\pi_t^1$ and $\pi_x^1$ vanish, there is no energy or momentum current flowing along the string.

We will now comment on the speed limit of these co-moving charges, which stems from the fact that the proper velocity $V$ of the string endpoints differs from the physical velocity $v$ in the four-dimensional field theory \cite{argyres}. From the metric (\ref{metric1}), we see that
\be
V=\fft{\cosh\rho_i}{\cosh (\rho_i-\rho_0)} v\,,
\ee
where $\rho_i=-\rho_1$ or $\rho_2$. In order to avoid a spacelike string worldsheet, we must have $V\le 1$. This corresponds to
\be
v\le v_{max}=\fft{\cosh (\rho_2-\rho_0)}{\cosh\rho_2}\,,
\ee
where we have taken $\rho_i=\rho_2$, since $\rho_2$ is closer to $\rho_0\ge 0$ than is $-\rho_1$. As is the case for steadily-moving strings in AdS black hole backgrounds, the speed limit depends on the mass parameter associated to the charge. This same speed limit applies to the curved strings which we now discuss. We see that $v_{max}=v_{crit}=1$ for $\rho_0=0$. On the other hand, for $\rho_0>0$, we find that $v_{max}<v_{crit}<1$.

\subsection{Curved strings}

We will now consider steadily-moving curved strings, which are described by solutions of the form $x(t,\rho)=x(\rho)+vt$. The term with the time derivative in (\ref{eom}) vanishes and we are left with
\be\label{eom2}
\fft{\partial}{\partial\rho} \Big( \fft{\cosh^2(\rho-\rho_0) \cosh^2\rho\ x^{\prime}}{\sqrt{-g}}\Big)=0\,.
\ee
where
\be
-\fft{g}{\ell^4}=\cosh^2(\rho-\rho_0)+\cosh^2(\rho-\rho_0) \cosh^2\rho\ x^{\prime 2}-v^2 \cosh^2\rho\,.
\ee
The first integral of (\ref{eom2}) is
\be
\fft{\cosh^2(\rho-\rho_0) \cosh^2\rho\ x^{\prime}}{\sqrt{-g/\ell^4}}=C\,.
\ee
where $C$ is an integration constant. Thus,
\be\label{xeqn}
x^{\prime 2}=\fft{C^2}{\cosh^2(\rho-\rho_0)\cosh^2\rho}\ \Big( \fft{\cosh^2(\rho-\rho_0)-v^2 \cosh^2\rho}{\cosh^2(\rho-\rho_0)\cosh^2\rho-C^2}\Big)\,.
\ee
Solving for $-g$ gives
\be
-\fft{g}{\ell^4}=\cosh^2(\rho-\rho_0)\cosh^2\rho\ \Big( \fft{\cosh^2(\rho-\rho_0)-v^2\cosh^2\rho}{\cosh^2(\rho-\rho_0)\cosh^2\rho-C^2}\Big)\,.
\ee
For $v\le v_{crit}$ (with the exception of $\rho_0=0$ and $v=1$) then $-g>0$ everywhere along the string provided that $|C|<C_{crit}$ where
\be\label{Ccrit}
C_{crit}=\fft12\big( \cosh (2\mbox{min}[\rho_2,\rho_0/2]-\rho_0)+\cosh\rho_0\big)\,.
\ee
In the limit that $\rho_0=0$, solutions exist with $|C|<1$ for all $v<1$. However, for $\rho_0=0$ and $v=1$, the induced metric is degenerate everywhere, regardless of the value of $C$. Strings with $v>v_{crit}$ can still have $-g>0$ everywhere provided that $\rho_2<\rho_{crit}$ and $|C|<C_{crit}$, where $\rho_{crit}$ and $C_{crit}$ are given by (\ref{pcrit}) and (\ref{Ccrit}), respectively. From the field theory perspective, this corresponds to an upper bound on the mass of the type 2 charge. For $v>v_{crit}$, one can also have a string that lives entirely in $\rho>\rho_{crit}$, provided that $|C|>C_{crit}$. This corresponds to a light-heavy pair of type 2 charges with a lower bound on their masses.

The distance between the charges is given by
\be
L=C\int_{-\rho_1}^{\rho_2} \fft{d\rho}{\cosh (\rho-\rho_0)\cosh\rho} \sqrt{\fft{\cosh^2 (\rho-\rho_0)-v^2\cosh^2\rho}{\cosh^2 (\rho-\rho_0)\cosh^2\rho-C^2}}\,,
\ee
For cases in which we require that $|C|<C_{crit}$, there is an upper bound on $L$. However, if the endpoints of the string are allowed to move freely, then the string will tend to straighten out to the $C=0$ configuration so that it minimizes its energy. On the other hand, for the string corresponding to a light-heavy pair of type 2 charges moving with $v>v_{crit}$, we must have $|C|>C_{crit}$ and there is a corresponding lower bound on $L$. This string cannot straighten out without decreasing its speed.

From (\ref{currents}), we find that the rate at which energy and momentum flow through the string can be written in terms of the constant $C$ as
\be
\pi_t^1=\fft{\sqrt{\lambda}}{2\pi} Cv\,,\qquad -\pi_x^1=\fft{\sqrt{\lambda}}{2\pi}C\,,
\ee
where we have used $T_0=\sqrt{\lambda}/(2\pi \ell^2)$. Thus, for nonzero C, energy and momentum are transferred from the leading charge to the lagging one. If there is no external force present, then the leading charge presumably slows down and the lagging charge speeds up until an equilibrium state is reached corresponding to a steadily-moving straight string.

\section{Strings on another Einstein-Gauss-Bonnet wormhole}

Thus far in this paper, we have discussed strings moving on Einstein-Gauss-Bonnet wormholes which have two asymptotically locally AdS regions. We now turn to another wormhole solution which has the metric \cite{troncoso2}
\be\label{altmetric}
ds^2=\ell^2 (-e^{2\rho} dt^2+d\rho^2+\cosh^2\rho\ d\Sigma_3^2)\,.
\ee
The geometry is locally AdS for $\rho\rightarrow+\infty$. The asymptotic geometry for $\rho\rightarrow -\infty$ is also smooth and can be roughly thought of as a five-dimensional Lorentzian analog of the Sol manifold \cite{thurston} .

As before, we will work in the static gauge $(\sigma, \tau)=(\rho,t)$. The Nambu-Goto action on this background gives rise to the equation of motion
\be
\fft{\partial}{\partial\rho} \Big( \fft{e^{2\rho} \cosh^2\rho\ x^{\prime}}{\sqrt{-g}}\Big)-\cosh^2\rho\ \fft{\partial}{\partial t} \Big( \fft{\dot x}{\sqrt{-g}}\Big)=0\,.
\ee
where
\be\label{altg}
-\fft{g}{\ell^4}=e^{2\rho}+e^{2\rho} \cosh^2\rho\ x^{\prime 2}-\cosh^2\rho\ \dot x^2\,.
\ee

A straight static string extending from $\rho=\rho_2$ to $\rho=-\rho_1$ has an energy
\be\label{altstraightE}
E=T_0\ell^2 (\mbox{e}^{\rho_2}-\mbox{e}^{-\rho_1})\,.
\ee
Expressing the equation of motion in the alternative gauge $(\sigma,\tau)=(x,t)$, it can be shown that there are also straight string solutions at constant $\rho=0$.

We will now discuss some properties of static curved strings on this background. As was the case with the previous class of wormholes, curved strings can have both endpoints on the same side of this wormhole or on opposite sides. In the former case, a string that is far from the neck of the wormhole has an energy $E\sim 1/L$, since this is the UV conformal regime of the theory. In this case, the pair of charges of the same type exhibit a Coulomb potential. As the string dips closer to the wormhole, IR effects show up as terms in the energy that are of higher order in $L$. There is a one-to-one correspondence between $L$ and the location of the turning point $\rho_t$. Thus, the charges do not exhibit a screening length. 

For a curved string with endpoints on opposite sides of the wormhole,
\be
L=C \int_{-\rho_1}^{\rho_2} \fft{d\rho}{\cosh\rho \sqrt{e^{2\rho} \cosh^2\rho-C^2}}\,.
\ee
For small separation $L$, this can be expanded in $C\ll 1$ so that
\be\label{altLC}
L\approx C\left[ \mbox{sech}\rho_2-\mbox{sech}\rho_1+2\arctan \left( \fft{e^{\rho_1+\rho_2}-1}{e^{\rho_1}+e^{\rho_2}}\right)\right]\,.
\ee
Euclideanizing the metric in order to compute the energy of the string configuration in terms of its action, we find the energy of a pair of charges to be
\be
E=T_0\ell^2 \int_{-\rho_1}^{\rho_2} d\rho\ \fft{e^{2\rho}\cosh\rho}{\sqrt{e^{2\rho}\cosh^2\rho-C^2}}\,.
\ee
Expanding $E$ in the small separation limit $C\ll 1$ and using (\ref{altLC}) to express $C$ in terms of $L$, we find
\be
E=E_{straight}+\fft12 k L^2\,,
\ee
where $E_{straight}$ is the energy of a straight string given by (\ref{altstraightE}) and
\be
k=\fft{\sqrt{\lambda}}{2\pi} \left[ \mbox{sech}\rho_2-\mbox{sech}\rho_1+2\arctan \left( \fft{e^{\rho_1+\rho_2}-1}{e^{\rho_1}+e^{\rho_2}}\right)\right]^{-1}\,.
\ee
Thus, for small $L$, a pair of charges of opposite types are confined and exhibit a spring-like potential. 
It can be shown that $k$ decreases monotonically with the mass parameters, and therefore heavy charges are less sensitive to the confining potential.

We will now consider steadily-moving straight strings which, from (\ref{altg}) with $\dot x=v$ and $x^{\prime}=0$, we find can exist in the region $\rho>\rho_{crit}$, where
\be\label{altpcrit}
\rho_{crit}=\ln \sqrt{\fft{v}{2-v}}\,.
\ee
For $v\rightarrow 0$, $\rho_{crit}\rightarrow -\infty$ while for $v\rightarrow 1$, $\rho_{crit}\rightarrow 0$. For fixed $\rho_1$, we can have steadily-moving straight strings only for $\rho_{crit}<-\rho_1$, which corresponds to $v<v_{crit}$, where
\be
v_{crit}=\fft{2}{1+\mbox{e}^{2\rho_1}}\,.
\ee
Interestingly enough, the critical velocity depends on the mass parameter of the lagging charge, which in this case is the type 1 charge. If $\rho_1=0$ then $v_{crit}=1$, and $v_{crit}$ monotonically decreases with the mass parameter for $\rho_1\neq 0$. The energy and momentum of such a string are given by
\bea
E &=& T_0\ell^2 \int_{-\rho_1}^{\rho_2} d\rho\ \fft{\mbox{e}^{2\rho}}{\sqrt{\mbox{e}^{2\rho}-v^2\cosh^2\rho}}\,,\nn\\
p &=& T_0\ell^2v\int_{-\rho_1}^{\rho_2} d\rho\ \fft{\cosh^2\rho}{\sqrt{\mbox{e}^{2\rho}-v^2\cosh^2\rho}}\,.
\eea

We will now consider steadily-moving curved strings. For $x(t,\rho)=x(\rho)+vt$, we get
\be
x^{\prime 2}=\fft{C^2}{e^{2\rho} \cosh^2\rho} \Big( \fft{e^{2\rho}-v^2 \cosh^2\rho}{e^{2\rho} \cosh^2\rho-C^2}\Big)\,.
\ee
Solving for $-g$ yields
\be\label{galt2}
-\fft{g}{\ell^4}=\mbox{e}^{2\rho} \cosh^2\rho \Big( \fft{e^{2\rho}-v^2\cosh^2\rho}{e^{2\rho} \cosh^2\rho-C^2}\Big)\,.
\ee
Strings can exist with $-\rho_1>\rho_{crit}$, where $\rho_{crit}$ is given by (\ref{altpcrit}), provided that $C<C_{crit}$, where
\be
C_{crit}=\ft12 \left( e^{-2\rho_1}+1\right)\,.
\ee
Thus, along with the upper bound on the mass parameter associated with the type 1 charge, there is an upper bound on the parameter $C$. This latter restriction corresponds to an upper bound on the distance between the charges
\be
L<C_{crit} \int_{-\rho_1}^{\rho_2} \fft{d\rho}{e^{\rho} \cosh\rho} \sqrt{\fft{e^{2\rho}-v^2\cosh^2\rho}{e^{2\rho}\cosh^2\rho-C_{crit}^2}}\,,
\ee
along with the following upper bounds on the rate of energy and momentum flow between the charges:
\be
\pi_t^1<\fft{\sqrt{\lambda}\ v}{4\pi} \left( \mbox{e}^{-2\rho_1}+1\right)\,,\qquad -\pi_x^1<\fft{\sqrt{\lambda}}{4\pi} \left( \mbox{e}^{-2\rho_1}+1\right)\,.
\ee
There can also be strings in the region $\rho<\rho_{crit}$, which correspond to light-heavy pairs of type 1 charges. In this case, $\rho_{crit}$ corresponds to a lower bound on the associated mass parameters, and $|C|>C_{crit}$ corresponds to a lower bound on distance between the charges as well as the rate of energy and momentum transfer.

Since the numerator and denominator in the expression (\ref{galt2}) are both positive for large and positive $\rho$, it is possible for $-g>0$ even for strings passing through the critical point $\rho_{crit}$, provided that the numerator and denominator both change sign there. This corresponds to the case in which $v\ge v_{crit}$. Then no restrictions are placed on the mass parameters but we must have
\be
C=\fft{1}{2-v}\,.
\ee
This specifies the distance between the charges to be 
\be
L=\int_{-\rho_1}^{\rho_2} \fft{d\rho}{e^{\rho} \cosh\rho} \sqrt{\fft{v e^{-2\rho}+v+2}{(2-v)e^{2\rho}+4-v}}\,.
\ee
We have confirmed numerically that $L$ increases monotonically with $v$ for $v\ge v_{crit}$. The rate of the energy and momentum flow along the string is given by
\be
\pi_t^1=\fft{\sqrt{\lambda}\ v}{2\pi (2-v)}\qquad -\pi_x^1=\fft{\sqrt{\lambda}}{2\pi (2-v)}\,.
\ee

\begin{figure}[ht]
   \epsfxsize=2.6in \centerline{\epsffile{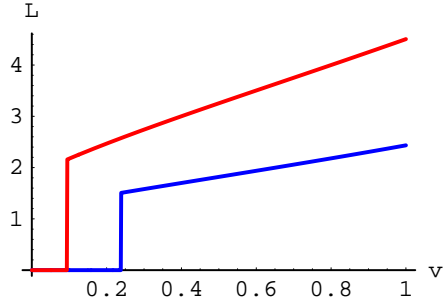}}
   \caption[FIG. \arabic{figure}.]{\footnotesize{The separation gap $L$ as a function of velocity $v$ for $\rho_1=\rho_2=1$ (blue) and $\rho_1=1.5$, $\rho_2=1$ (red).}}
\label{fig11}
\end{figure}

From the field theory perspective, we have the following picture. There is no interaction between two coincident charges of opposite types co-moving with a speed of $v<v_{crit}$. However, for $v=v_{crit}$, the charges exhibit a separation gap, which increases with larger $v$. This is shown in Figure \ref{fig11} for $\rho_1=\rho_2=1$ (blue line) for which $v_{crit}\approx 0.0238$, and $\rho_1=1.5$ and $\rho_2=1$ (red line) for which $v_{crit}\approx 0.095$. 
For a given $v$ above the critical value, the separation gap $L$ increases with $\rho_1$ and is less sensitive to the value of $\rho_2$. Also, for $v<v_{crit}$, no energy or momentum is transferred between the charges. However, for $v=v_{crit}$,
\be
\pi_t^1=\fft{\sqrt{\lambda}}{2\pi}\ \mbox{e}^{-2\rho_1}\,,\qquad -\pi_x^1=\fft{\sqrt{\lambda}}{4\pi} \left( \mbox{e}^{-2\rho_1}+1\right)\,.
\ee
which both {\it decrease} with the mass parameter of the lagging charge. This is certainly unexpected from the field theory side. However, on the gravity theory side, this arises as a peculiar property of the wormhole background. 

\begin{figure}[h]
\begin{center}
$\begin{array}{c@{\hspace{1.25in}}c}
\epsfxsize=1.7in \epsffile{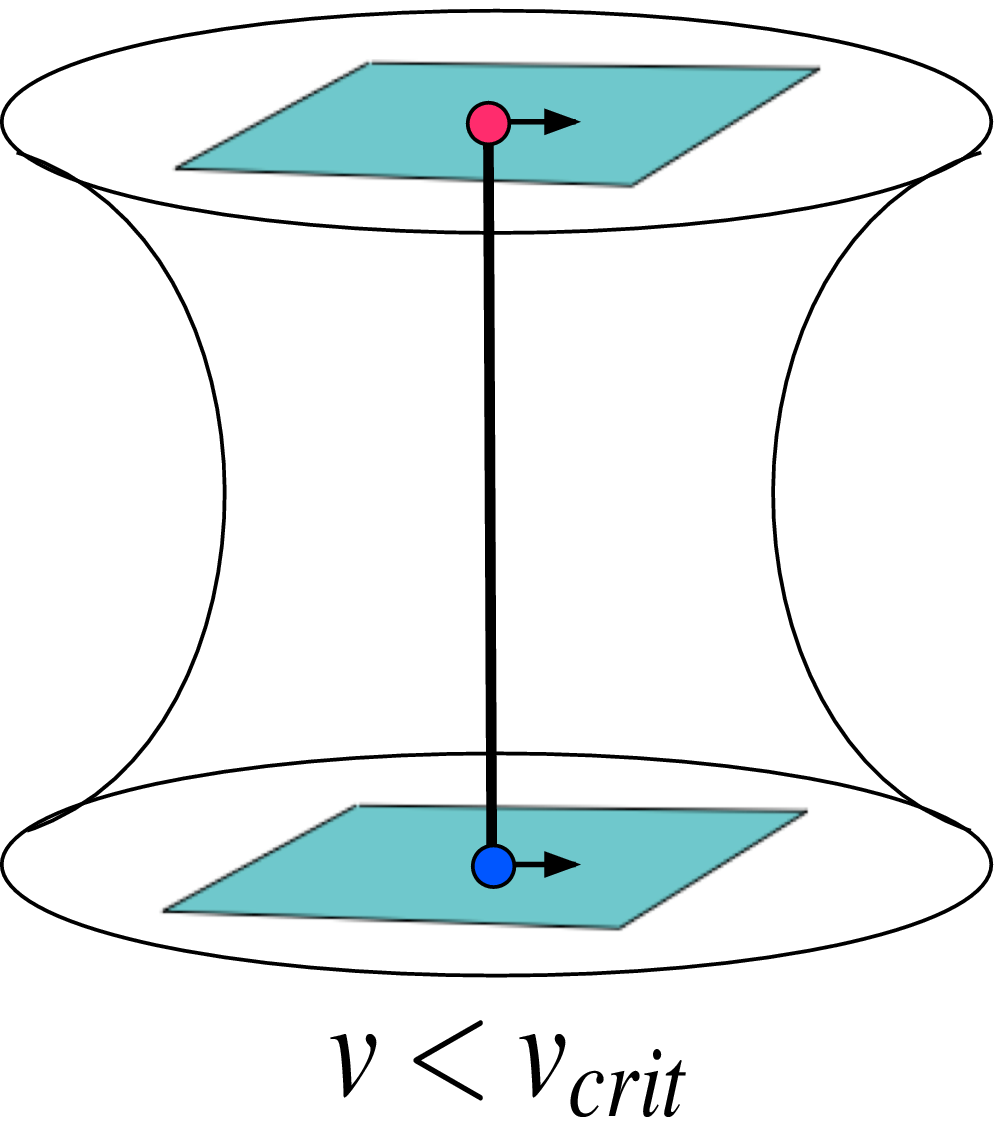} &
\epsfxsize=1.7in \epsffile{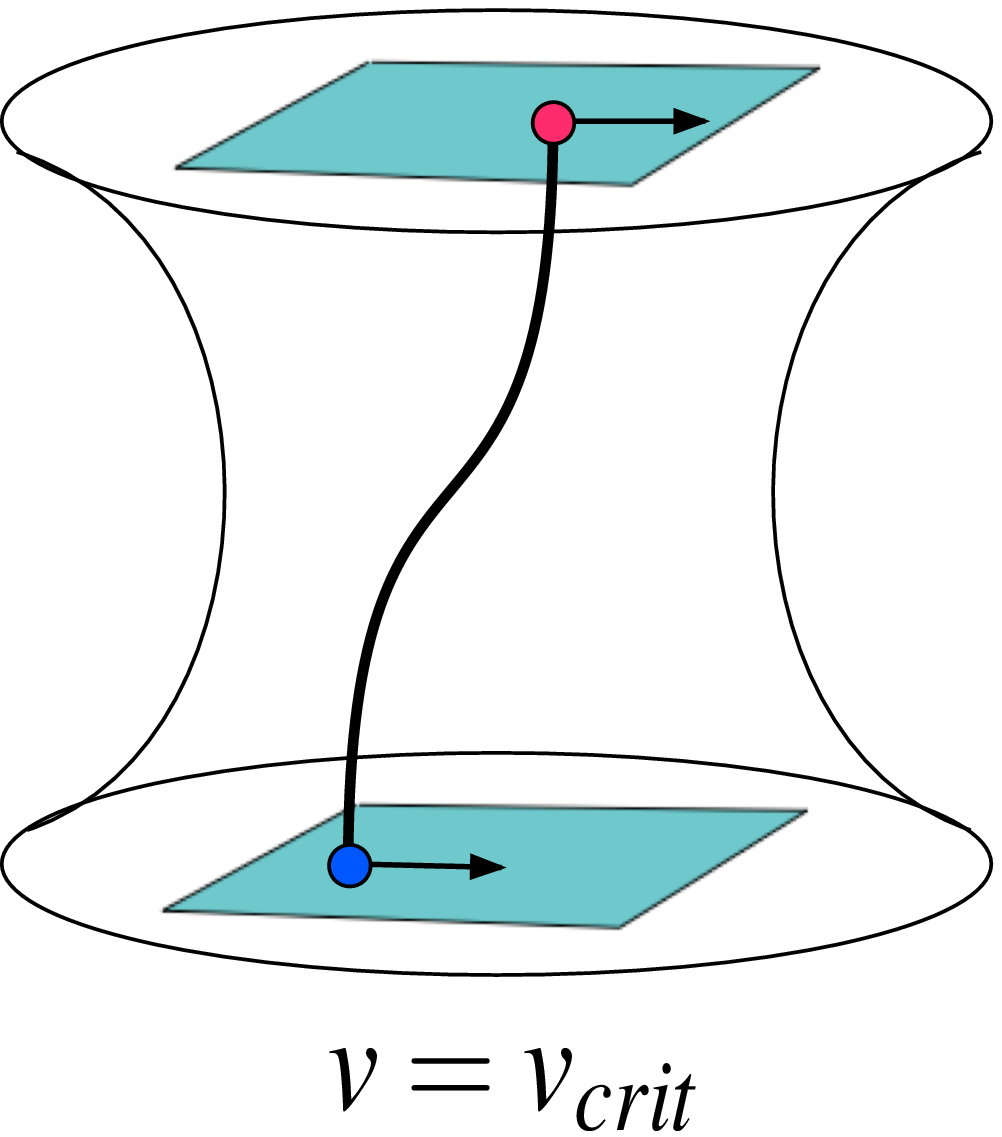}
\end{array}$
\end{center}
\caption[FIG. \arabic{figure}.]{\footnotesize{A steadily-moving string can stretch straight through the wormhole for $v<v_{crit}$, as shown in the left plot. However, a steadily-moving string with $v=v_{crit}$ must curve in order to be physically acceptable, as is shown in the right plot. Strings with $v>v_{crit}$ are spacelike.}}
   \label{fig2}
\end{figure}

As in the previous section, in order to avoid a spacelike string, we must have the proper velocity $V\le 1$. From the metric (\ref{altmetric}), this translates into $v\le v_{crit}$. Thus, even though $-g>0$ for strings which travel at $v>v_{crit}$ for an appropriate value of $C$, it turns out that such strings would still be spacelike. Perhaps if a string attempts to move faster than this speed limit, then the increased amount of energy and momentum transferred along the string, as shown in Figure \ref{fig11}, would immediately cause it to slow down. However, it would appear that it is possible for strings to travel at the critical speed itself, so long as there is a constant input of energy and momentum. The situations for $v<v_{crit}$ and $v=v_{crit}$ are depicted in Figure \ref{fig2}.

\section{Strings on AdS wormholes in supergravity}

Thus far, we have considered AdS wormholes which are solutions to Einstein-Gauss-Bonnet theory. One could ask which properties of strings on such backgrounds might simply be a reflection of the specific higher-derivative curvature terms in the theory. We will now consider a couple of examples of AdS wormholes which can be embedded within string theory. The similar behavior of strings for these cases suggests that strings on AdS wormholes share some universal features.

\subsection{A multi-boundary orbifold of AdS$_3$}

We will first consider the behavior of strings on a multi-boundary orbifold of AdS$_3$ \cite{balasubramanian}. The metric is given by
\be\label{metricads3}
ds_3^2=\ell^2\left( -dt^2+dx^2-2\sinh(2\rho) dtdx+d\rho^2\right)\,.
\ee
In the static gauge $(\sigma,\tau)=(\rho,t)$, the Nambu-Goto action on this background gives the equation of motion
\be
\fft{\partial}{\partial\rho} \left( \fft{\cosh^2(2\rho) x^{\prime}}{\sqrt{-g}}\right)-\fft{\partial}{\partial t}\left( \fft{\dot x-\sinh(2\rho)}{\sqrt{-g}}\right)=0\,,
\ee
where
\be
-\fft{g}{\ell^4}=1-\dot x^2+2\sinh(2\rho)\dot x+x^{\prime 2}\cosh^2(2\rho)\,.
\ee

From (\ref{currents}) and (\ref{generalEp}), we find that the energy of a straight string extended from $\rho=-\rho_1$ to $\rho_2$ is 
\be\label{straightEads3}
E=T_0\ell^2 (\rho_1+\rho_2)\,.
\ee
If one expresses the equation of motion in the gauge $(\sigma,\tau)=(x,t)$, then it can be shown that there are also straight strings lying along constant $\rho$ for any value of $\rho$. 

We will now consider a string with both endpoints on the same side of the wormhole at $\rho=\rho_2$. Since the geometry is asymptotically locally AdS, the endpoints of a string that is far from the wormhole exhibit Coulomb behavior. The distance $L$ between the endpoints is given by
\be
L=2\cosh(2\rho_t) \int_{\rho_t}^{\rho_2} \fft{d\rho}{\cosh(2\rho) \sqrt{\cosh^2(2\rho)-\cosh^2(2\rho_t)}}\,,
\ee
where the turning point $\rho_t$ can lie within the region $\rho_2>\rho_t\ge 0$. Thus, all strings bend towards the wormhole and none of them pass through the neck of the wormhole.

For a string with its endpoints on opposite sides of the wormhole, expanding in $C$ for small $L$ as in previous sections, we find
\be
E_{straight}+\fft12 k L^2\,,
\ee
where $E_{straight}$ is the energy of a straight string given by (\ref{straightEads3}) and
\be
k=\fft{\sqrt{\lambda}}{\pi \left[ \tanh \left( \fft{\rho_1}{2}\right)+\tanh \left( \fft{\rho_2}{2}\right)\right]}\,.
\ee
While $k$ decreases monotonically with the mass parameters, it asymptotically approaches the minimum value
\be
k_{min}=\fft{\sqrt{\lambda}}{2\pi}\,,
\ee
as the $\rho_i\rightarrow\infty$. 

As in the previous examples, when there is a pair of type 1 charges and a pair of type 2 charges there can be a transition. Namely, the charges of the same type exhibit Coulomb interaction for small $L$ and the opposite type charges do not interact, whereas for large $L$ the opposite type charges are confined and the charges of the same type are effectively screened.

We now turn to steadily-moving straight strings, for which 
\be
-\fft{g}{\ell^4}=1-v^2+2v \sinh(2\rho)\,.
\ee
In order for $-g>0$ everywhere along the string, it must lie in the region bounded by $\rho>\rho_{crit}$, where
\be\label{pcritads3}
\rho_{crit}=\fft12 \ln\ v\,.
\ee
For $v\rightarrow 0$, $\rho_{crit}\rightarrow -\infty$ while for $v\rightarrow 1$, $\rho_{crit}\rightarrow 0$. For fixed $\rho_1$, we can have steadily-moving straight strings only for $\rho_{crit}<-\rho_1$, which corresponds to
\be
v<v_{crit}=e^{-2\rho_1}\,.
\ee
The critical velocity depends on the mass parameter associated with the lagging charge, which in this case is the type 1 charge. If $\rho_1=0$, then $v_{crit}=1$. For fixed 't Hooft coupling, $v_{crit}$ decreases with the mass parameter and vanishes for infinite $\rho_1$.

We will now consider steadily-moving curved strings. For $x(t,\rho)=x(\rho)+vt$, we get
\be
x^{\prime 2}=\fft{C^2 \left( 1-v^2+2v\sinh(2\rho)\right)}{\cosh^2(2\rho) \left( \cosh^2(2\rho)-C^2\right)}\,,
\ee
with
\be
-\fft{g}{\ell^4}=\fft{\cosh^2(2\rho)\left( 1-v^2+2v\sinh(2\rho)\right)}{\cosh^2(2\rho)-C^2}\,.
\ee
Strings can exist with $-\rho_1>\rho_{crit}$, where $\rho_{crit}$ is given by (\ref{pcritads3}), provided that $C^2<1$. This bound on $C$ results in 
\be
L<\left[\fft{v}{2} \arctan \left( \fft{\alpha}{v-\sinh (2\rho)}\right) -\gamma^{-1} \arctan \left( \gamma\alpha\right) +\fft14 \ln \left( \fft{v^2+(\alpha+1)^2}{v^2+(\alpha-1)^2}\right)\right]\Big|_{-\rho_1}^{\rho_2}\,,
\ee
where
\be
\alpha=\sqrt{\gamma^{-2}+2v\sinh (2\rho)}\,,\qquad \gamma=\fft{1}{\sqrt{1-v^2}}\,,
\ee
along with the following bounds on the rate of energy and momentum flow between the charges:
\be
\pi_t^1<\fft{\sqrt{\lambda}\ v}{2\pi}\,,\qquad -\pi_x^1<\fft{\sqrt{\lambda}}{2\pi}\,.
\ee
There can also be strings in the region $\rho<\rho_{crit}$, which correspond to light-heavy pairs of type 1 charges. Then there is a lower bound on the associated mass parameters. Also, $C^2>1$, which corresponds to a lower bound on $L$ as well as the rate of energy and momentum transfer.

It is possible for $-g>0$ even for strings with $v\ge v_{crit}$ which pass through the critical radius $\rho_{crit}$ so long as the numerator and denominator of $-g$ both change sign there. While there are now no restrictions on the mass parameters, we must have 
\be
C=\fft{1+v^2}{2v}\,.
\ee
This specifies the distance between the charges to be
\be
L=\mbox{Im} \left[\fft{1}{i+v}\ \mbox{arctanh} \left( \fft{\sqrt{v^2+2v\sinh (2\rho)-1}}{i+v}\right)\right]\Big|_{-\rho_1}^{\rho_2}\,.
\ee
The rate of energy and momentum flow from the leading charge to the lagging one is given by
\be
\pi_t^1=\fft{\sqrt{\lambda}}{4\pi}\ (1+v^2)\,,\qquad -\pi_x^1=\fft{\sqrt{\lambda}}{4\pi v}\ (1+v^2)\,.
\ee

For $v<v_{crit}$, $L=0$. However, for $v=v_{crit}$, $L$ becomes
\be
L=\mbox{Im} \left[ \fft{1}{i+e^{-2\rho_1}}\ \mbox{arctanh} \left( \fft{\sqrt{e^{-4\rho_1}+2e^{-2\rho_1} \sinh (2\rho_2)-1}}{i+e^{-2\rho_1}}\right)\right]\,.
\ee
Likewise, no energy or momentum is transferred from one charge to the other for $v<v_{crit}$. However, for $v=v_{crit}$,
\be
\pi_t^1=\fft{\sqrt{\lambda}}{4\pi}\ \left(1+e^{-4\rho_1}\right)\,,\qquad -\pi_x^1=\fft{\sqrt{\lambda}}{2\pi} \cosh (2\rho_1)\,.
\ee
Curiously enough, the rate at which energy is transferred decreases with the mass parameter $\rho_1$ of the lagging charge, whereas the rate at which momentum is transferred increases with $\rho_1$.

From the metric (\ref{metricads3}), the proper velocity of the string endpoint at $\rho=-\rho_1$ is given by
\be
V=\sqrt{v^2+2v\sinh (2\rho_1)}\,,
\ee
where $v$ is the physical velocity in the dual two-dimensional field theory. In order to avoid a spacelike string, $V\le 1$, which corresponds to having $v\le v_{crit}$. Thus, it turns out that $v_{crit}$ is the speed limit of the pair of charges.

\subsection{The Maoz-Maldacena wormhole}

We will now consider some static properties of strings on a Euclidean AdS wormhole solution arising in five-dimensional gauged supergravity \cite{maoz}. The metric is given by
\be
ds_5^2=d\rho^2+e^{2\omega} \left( d\theta^2+\ft14 \sin^2\theta\ w^a w^a\right)\,,
\ee
where 
\be
e^{2\omega}=\fft{\sqrt{5}}{2}\cosh 2\rho-\fft12\,,
\ee
and $w^a$ are the left-invariant one-forms on $S^3$. There are also $SO(6)$ gauge fields given by
\be
A_{\mu}^{IJ}=i A_{\mu}^a L^{a IJ}+i \td A_{\mu}^a \td L^{a IJ}\,,
\ee
where $L^a$ and $\td L^a$ are generators of $SO(3)\times SO(3)$ and 
\be
A^a=\cos^2\fft{\theta}{2}w^a\,,\qquad \td A^a=\sin^2\fft{\theta}{2}w^a\,,
\ee
are an instanton and an anti-instanton which are $SO(5)$ symmetric under rotations of $S^4$. The corresponding gauge theory is ${\cal N}=4$ super Yang-Mills theory with an external fixed gauge field coupled to the $SO(6)$ currents. It was shown explicitly that the supergravity background is rather stable, and the field theory on each boundary is well-defined \cite{maoz}.

The action of the Euclidean string worldsheet is given by (\ref{EuclideanAction}). We will consider a string whose worldsheet lies along two of the $S^3$ directions, which we will refer to as $t$ and $x$. Since the string lies at a point in the $\theta$ direction, we will rescale the $x$ and $t$ coordinates to include constant factors in the metric. In the gauge $\sigma=\rho$ and $\tau=t$, we find that the equation of motion is given by
\be\label{mmeom1}
\fft{\partial}{\partial\rho}\left( \fft{e^{4\omega} x^{\prime}}{\sqrt{g}}\right)+e^{2\omega} \fft{\partial}{\partial t}\left( \fft{\dot x}{\sqrt{g}}\right) =0\,,
\ee
where
\be
g=e^{2\omega}\left( 1+\dot x^2+e^{2\omega} x^{\prime 2}\right)\,.
\ee
From (\ref{currents}) and (\ref{generalEp}), we find that the energy of a straight string extended from $\rho=-\rho_1$ to $\rho_2$ is 
\be\label{mmE}
E=-i \sqrt{\fft{\sqrt{5}-1}{2}} T_0\left( \mbox{EllipticE}\left[ i\rho_1,\fft{2\sqrt{5}}{\sqrt{5}-1}\right]+\mbox{EllipticE}\left[ i\rho_2,\fft{2\sqrt{5}}{\sqrt{5}-1}\right]\right)\,,
\ee
where EllipticE is the elliptic integral of the second kind. $E$ is a real function and increases monotonically with $\rho_i$.

In the gauge $\sigma =x$ and $\tau=t$, the equation of motion is
\be\label{mmeom2}
\fft{\partial}{\partial x}\left( \fft{e^{2\omega} \rho^{\prime}}{\sqrt{g}}\right) +e^{2\omega} \fft{\partial}{\partial t}\left( \fft{\dot\rho}{\sqrt{g}}\right) =\fft{\sqrt{5}\sinh 2\rho}{2\sqrt{g}} \left( \sqrt{5}\cosh 2\rho-1+\rho^{\prime 2}+\dot\rho^2\right)\,,
\ee
where
\be
g=e^{2\omega} \left( e^{2\omega}+\rho^{\prime 2}+\dot\rho^2\right)\,.
\ee
We see that there is also a straight string solution at constant $\rho=0$.

We will now consider string solutions with both endpoints on the same side of the wormhole at $\rho=\rho_2$. Since the geometry is asymptotically hyperbolic, the endpoints of a string that is far from the wormhole exhibit Coulomb behavior. As the distance $L$ between the endpoints is increased, the turning point $\rho_t$ gets closer to the wormhole. From (\ref{mmeom2}), we find that
\be
L=2e^{2\omega_t} \int_{\rho_t}^{\rho_2} \fft{d\rho}{e^{\omega} \sqrt{e^{4\omega}-e^{4\omega_t}}}\,,
\ee
where $\omega_t=\omega|_{\rho=\rho_t}$. By analyzing the above formula, we see that there can be string solutions with $\rho_2>\rho_t\ge 0$. It seems that all strings bend toward the wormhole and none of them can pass through the neck of the wormhole, as opposed to the wormhole background discussed in section 3.

For a string with its endpoints on opposite sides of the wormhole,
\be
L=C \int_{-\rho_1}^{\rho_2} \fft{d\rho}{e^{\omega} \sqrt{e^{4\omega}-C^2}}\,.
\ee
For small separation $L$, we can expand in $C\ll 1$ so that
\be
L\approx C \int_{-\rho_1}^{\rho_2} d\rho\ e^{-3\omega}\,.
\ee
From the worldsheet action given by (\ref{EuclideanAction}), we find the energy of the string to be
\be
E=T_0 \int_{-\rho_1}^{\rho_2} d\rho \fft{e^{3\omega}}{\sqrt{e^{4\omega}-C^2}}\,.
\ee
Expanding $E$ in small $C$ and then expressing $C$ in terms of $L$, we find
\be
E\approx E_{straight}+\fft12 k L^2\,,
\ee
where $E_{straight}$ is the energy of a straight string which is given by (\ref{mmE}) and
\be
k=T_0 \left[ \int_{-\rho_1}^{\rho_2} d\rho\ e^{-3\omega}\right]^{-1}\,.
\ee
Thus, for small $L$, a pair of charges of opposite types are confined by a spring-like potential. It can be shown that $k$ decreases monotonically with the mass parameters, which means that heavy charges are less sensitive to the confining potential.

When there is a pair of type 1 charges a distance $L$ apart and a pair of type 2 charges also a distance $L$ apart, the same types of transitions can occur as with the previously-discussed wormholes. Namely, for small $L$ only charges of the same type interact with each other, whereas for large $L$ only opposite type charges interact with each other. This was discussed in a bit more detail in section 3.2 and illustrated in Figure \ref{figtrans}.

\section*{Acknowledgments}

We would like to thank Philip Argyres, Mohammad Edalati, Hong L\"u and Juan Maldacena for useful correspondence and conversations. M.A., F.R. and C.S.-V. are supported in part by the Emerging Scholars Program. F.R. and C.S.-V. are supported in part by New York City Louis Stokes Alliance for Minority Participation in Science, Mathematics, Engineering and Technology.


\end{document}